\documentclass[pre,twocolumn,floatfix,superscriptaddress,showpacs]{revtex4}
\usepackage{graphicx,amsmath,amssymb,mathptm}
\begin{document}
\title{Growing network model for community with group structure}
\author{Jae Dong Noh}
\affiliation{Department of Physics, Chungnam National University, Daejeon
305-764, Korea}
\author{Hyeong-Chai Jeong}
\affiliation{Department of Physics and IFP, Sejong University,
Seoul 143-747, Korea}
\author{Yong-Yeol Ahn}
\affiliation{Department of Physics, Korea Advanced Institute of Science and
Technology, Daejeon 305-701, Korea}
\author{Hawoong Jeong}
\affiliation{Department of Physics, Korea Advanced Institute of Science and
Technology, Daejeon 305-701, Korea}
\date{\today}

\begin{abstract}
We propose a growing network model for a community with a group structure.  
The community consists of individual members and groups, gatherings of
members. The community grows as a new member is introduced 
by an existing member at each time step. 
The new member then creates a new group or joins one of the groups of the
introducer. 
We investigate the emerging community structure analytically
and numerically. The group size distribution shows a 
power law distribution for a variety of growth rules,
while the activity distribution follows an exponential or a power
law depending on the details of the growth rule. 
We also present an analysis of empirical data from 
on the online communities,
the ``Groups'' in \url{http://www.yahoo.com} and 
the ``Cafe'' in \url{http://www.daum.net},
which shows a power law distribution
for a wide range of group sizes. 
\end{abstract}

\maketitle

\section{Introduction}\label{sec:1}
Emergent properties of artificial or natural complex systems
attract growing interests recently. Some of them are conveniently
modeled with a network, where constituting ingredients and
interactions are represented with vertices and links, respectively.
Watts and Strogatz demonstrated that real-world networks display the
small-world effect and the clustering property, which cannot be explained
with the regular and random networks~\cite{Watts98}. 
Later on, in the study of the WWW network, Albert {\em et al.}
found that the degree, the number of attached links, of each 
vertex follows a power-law distribution~\cite{Albert99}. 
Those works trigger a burst of researches on 
the structure and the organization principle of complex
networks~(see Refs.\cite{Albert02,Dorogovtsev02,Newman03SIAM} for reviews).

Many real-world networks, e.g., in biological, social, and technological 
systems, are found to obey the power-law degree
distribution~\cite{Albert02}.
A network with the power-law distribution is called a scale-free~(SF) 
network. One of the possible mechanism for the power law
is successfully explained with the
Barab\'asi-Albert~(BA) model~\cite{Barabasi99}. The model assumes that 
a network is growing and that the rate acquiring a new link for 
an existing vertex 
is proportional to a popularity measured by its degree. The
popularity-based growth appears very natural since, e.g., creating a new web
site, one would link it preferentially to popular sites having many links.
With the BA and related network models, structural and dynamical properties
of networks have been explored extensively.

On the other hand, there exists another class 
of networks which have a group structure. 
Consider, for example, 
online communities such as the ``Groups''
operated by the Yahoo~(\url{http://www.yahoo.com}) 
and the ``Cafes'' operated by the 
Korean portal site Daum~(\url{http://www.daum.net}). 
They consist of individual members and groups, gatherings of
members with a common interest, and growth of the community 
is driven not only by
members but also by groups. A community evolves as an individual
registers as a new member. The new comers can create
new groups with existing members or joins existing groups.
The online community
is a rapidly growing social network~\cite{comment_daum}.
The emerging structure would be distinct from that observed in networks
without the group structure. 
In this paper, we propose a growing network model for the community with 
the group structure. We model the community with a bipartite network
consisting 
of two distinct kinds of vertices representing members and groups, 
respectively. A link may exist only between a member vertex and a group
vertex, which represents a membership relation. 

The bipartite network~\cite{Newman01a} 
has been considered in the study 
of the movie actor network~\cite{Watts98} consisting of actors and movies,
the scientific collaboration network~\cite{Newman01a,Goldstein04}
of scientists and articles, and the company director
network~\cite{Newman01a} of directors and boards of directors.
Usually those networks are treated as unipartite by projecting out one kind
of vertices of less interest~\cite{Newman01b,Newman03}. 
Some biological and social networks are known to have a modular
structure~\cite{Girvan02,Ravasz03}, where vertices in a common module are
densely connected while vertices in different modules are sparsely
connected. The modular structure is coded implicitly in the connectivity
between vertices. Unipartite network models with the modular structure 
were also studied in
Refs.~\cite{Ravasz03,Watts02,Motter03,Skyrms00,Jin01,Gronlund04,DHKim03}, 
where vertices form modules which in turn form bigger modules 
hierarchically~\cite{Watts02,Ravasz03,Motter03} or the modular structure
emerges dynamically as a result of social
interactions~\cite{Skyrms00,Jin01,Gronlund04,DHKim03}.
In Ref.~\cite{DHKim03}, each vertex is assigned to a 
Potts-spin-like variable pointing to its module~\cite{DHKim03}. 
These studies on the group structures
of networks have mainly focused on the groups with finite number 
of members. However, there are groups in the real-world online community 
which keep growing as the community evolves.

Reflecting growing dynamics of the real-world online community, our model
takes account of the group structure explicitly with a bipartite network
consisting of member and group vertices. 
Upon growing, both the member and group vertices evolve in time.
We study the dynamics of the size of groups and the activity 
of the members. The size of a group is defined as the number of
members in the group and the activity of a member is the 
number of groups in which the member participates.
When the community grows large enough, 
the group size distribution shows
a power law distribution unlike the network models 
studied previously~\cite{Watts02,DHKim03}.
To test our model, we analyze the empirical data from on the 
online communities,
the ``Groups'' in \url{http://www.yahoo.com} and 
the ``Cafe'' in \url{http://www.daum.net}
and show that both communities indeed
show power law group size distributions
for wide ranges of group sizes.

This paper is organized as follows. In Sec.~\ref{sec:2}, we introduce the
growing network model.
Depending on the choice of detailed dynamic rules, one may consider a few
variants of the model. Characteristics such as the group size distribution,
the member activity distribution, and the growth of the number of groups 
are studied analytically in a mean field theory and 
numerically in Sec.~\ref{sec:3}. 
Those characteristics are also calculated for the real-world
online communities and compared with the model results.
We conclude the paper with summary in Sec.~\ref{sec:4}.

\section{Model}\label{sec:2}
We introduce a model for a growing community with the group structure. 
The community grows by adding a new member at a time, who may open a new
group or join an existing group~\cite{comment_Simon}.
Following notations are adopted: A member entering the community 
at time step $i$ is denoted by $I_i$. The activity, the number of 
participating groups, of $I_i$ is denoted by $A_i$. 
As members enter the community, new groups are created or 
existing groups expand. The $\alpha$th group is denoted by $G_\alpha$, 
its creation time by $\tau_\alpha$,
and its size by $S_\alpha$. The total number of members and groups is 
denoted by $N$ and $M$, respectively.

Initially, at time $t=0$, the community is assumed to be inaugurated by 
$m_0$ members, denoted by  $I_{-(m_0-1)},\ldots,I_0$, belonging
to an initial group $G_1$. That is, we have that $N(t=0)=m_0$, $M(t=0)=1$,
$A_j(t=0)=1$ for $j=-(m_0-1),\cdots,0$, $\tau_1=0$, and $S_1(t=0)=m_0$. 
At time $t$, a new individual
$I_t$ is introduced into the community and becomes a member by repeating the
following procedures until its activity reaches $m$:
\begin{itemize}
\item {\textbf{Selection}} : It selects a partner $I_{j}$
among existing members $\{I_{k<t}\}$
with a selection probability $P^S_j$.
\item {\textbf{Creation or Joining}} : 
With a creation probability $P^C_j$, it creates a new group $G_{M+1}$
with the  
partner $I_j$. 
Otherwise, it selects randomly one of the
groups of $I_j$ with the equal probability and joins it.
If $I_t$ is already a member of the selected group, then
the procedure is canceled.
\end{itemize}

A specific feature of the model varies with the choice of those
probabilities $P^S$ and $P^C$. 
Regarding to the selection, 
simplest is the random choice among existing members
with the equal probability $P^S_j = 1/(m_0+t-1)$. 
Note that the selection may be regarded as an invitation of a new member
by existing members. Then, it may be natural to assume that active
members invite more newcomers. Such a case is modeled with a preferential
selection probability $P^S_j = A_j / (\sum_{k<t} A_k)$.
After selecting a partner $I_j$, the newcomer may create a new group or 
join one of $I_j$'s groups with the equal probability. 
In that case the creation 
probability is variable as $P^C_j = 1 / (A_j + 1 )$.
In the other case, it may create a new group with a fixed 
probability $P^C_j = \omega$. 
Combining the strategies in the two procedures,
we consider the possible four different growth models denoted by RV,
RF, PV, and PF, respectively. 
Here, R~(P) stands for the random~(preferential) selection,
and V~(F) for the group creation with the variable~(fixed)
probability.
For example, the RF model has the selection probability,
$P^S_j = 1/(m_0+t-1)$ and the creation probability, 
 $P^C_j = 1 / (A_j + 1 )$.
The growth rules are summarized in Table~\ref{table1}.

\begin{figure}[t]
\includegraphics*[width=0.7\columnwidth]{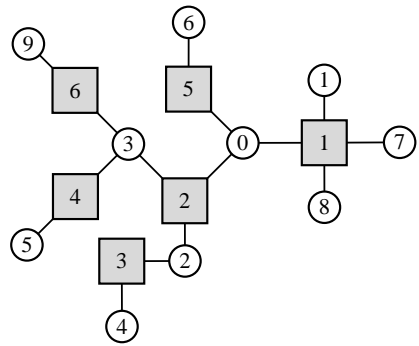}
\caption{A network for the RV model with $m_0=m =1$ and $N=10$ with 
six groups. The symbol
{\large\textcircled{{\small $i$}}} and 
\fbox{$\alpha$} 
represents
 a member $I_i$ and a group $G_\alpha$, 
respectively.}
\label{fig1}
\end{figure}

\begin{figure}[t]
\includegraphics*[width=\columnwidth]{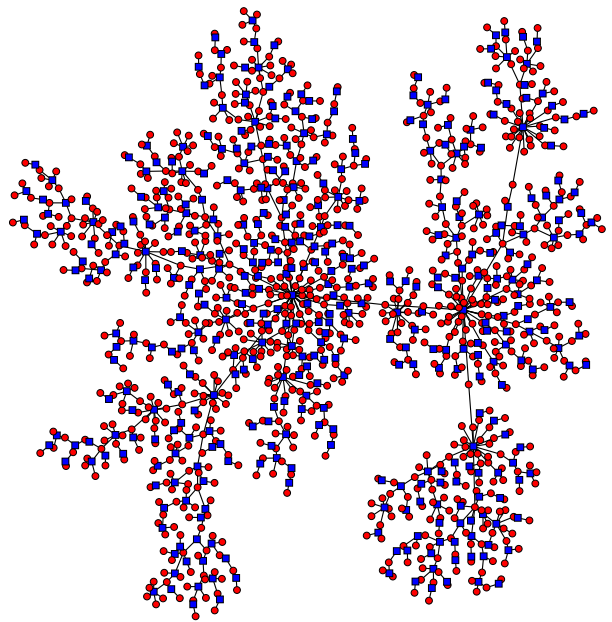}
\caption{ (color online) A network for the RV model with $m_0=m =1$ and 
$N=1000$. A square~(circle) symbol stands for a group~(member).}
\label{fig2}
\end{figure}

The whole structure of the community is conveniently represented with a
bipartite network of two kinds of vertices; one for the group and the other
for the member. A link exists only from a member vertex to a 
group vertex to which it belongs. The member activity and the group size
correspond to the degree of the corresponding vertex.
Figure~\ref{fig1} shows a typical 
network configuration for the RV model with $m_0=m=1$.
To help readers understand the growth dynamics, we add the indices for
members $I_i$ and groups $G_\alpha$ in the figure. 
It is easily read off that $I_1$ selects $I_0$ and becomes a member of 
$G_1$ at $t=1$ and that $I_2$ opens a new group $G_2$ with $I_0$ at $t=2$,
and so on. Figure~\ref{fig2} shows a configuration of a RV network with
$m=m_0=1$ grown up to $N=1000$ members with $M=452$ groups.
It is noteworthy that there appear hub groups having a lot of members.
The emerging structure of the network will be studied in the next section.

\begin{table}
\caption{Model description and mean field results for the 
group size distribution exponent $\gamma$. 
Here, $\Theta_{RV}$ and $\Theta_{PV}$ are the group number growth rate given
in Eqs.~(\ref{eq:Theta}) and (\ref{eq:Theta_PV}), respectively.
The activity distribution follows
a power law only for the PF model with the exponent
$\lambda=2+1/\omega$.
}
\label{table1}
\begin{ruledtabular}
\begin{tabular}{l|cc}
   &  R $\left(P_j^S = \frac{1}{m_0+t-1}\right)$  
   &  P $\left(P_j^S = \frac{A_j}{\sum_{k<t}A_k}\right)$ \\ \hline
V $\left(P_j^C=\frac{1}{A_j+1}\right)$ 
   & $1+\Theta_{RV}^{-1}$
   & $1 + \Theta_{PV}^{-1}$ \\
F $\left(P_j^C=\omega\right)$ 
   & $ 2 / (1-\omega)$
   & $ 2 / (1-\omega)$
\end{tabular}
\end{ruledtabular}
\end{table}

\section{Network Structure}\label{sec:3}
The number of groups $M(t)$, the activity of each member $A_i(t)$, and 
the size of each group $S_\alpha(t)$
increase as the network grows.
With those quantities, we characterize the growth dynamics and the network
structure. 
In the following, we study the dynamics of those quantities averaged over
network realizations. For simplicity's sake, we make use of the same
notations for the averaged quantities.
The network dynamics implies that they evolve in time as follows:
\begin{eqnarray}
A_i(t+1) &=& A_i(t) + m P^S_i P^C_i \label{delA} \\
M(t+1) &=& M(t) + m \sum_{j\leq t} P^S_j P^C_j \label{delM} \\
S_\alpha(t+1) &=& S_\alpha(t) + m \sum_{j\leq t} P^S_j \chi_{j\alpha}
(1-P^C_j)/A_j \ , \label{delS}
\end{eqnarray}
where $\chi_{j\alpha}=1$ if $I_j$ belongs to $G_\alpha$ or 0 otherwise.
The initial conditions are given by $A_i(t=i)=m$, $M(t=0)=1$, and
$S_\alpha(t=\tau_\alpha)=2$ with $\tau_\alpha$ the creation time of
$G_\alpha$. We analyze the equations in a continuum limit and 
in a mean field scheme, neglecting any correlation among dynamic variables. 

Firstly we consider the RV model. Using the corresponding $P^C$ and $P^S$
in Table~\ref{table1}, Eqs.~(\ref{delA},\ref{delM},\ref{delS}) become 
\begin{eqnarray}
\frac{dA_i}{dt} &=& \frac{m}{(A_i+1)(m_0+t)}  \\
\frac{dM}{dt} &=& \frac{1}{(m_0+t)} \sum_{j\leq t} \frac{m}{(A_j+1)} \\
\frac{dS_\alpha}{dt} &=&  \left(\frac{1 }{m_0+t}\right)
\left(\frac{S_\alpha}{m_0+t}\right) \sum_{j\leq t} \frac{m}{(A_j+1)} \ ,
\label{eq:dotS_RV}
\end{eqnarray}
where we approximate $\chi_{j\alpha}$ in Eq.~(\ref{delS}) with 
$\frac{S_\alpha}{(m_0+t)}$,
the fraction of members of $G_\alpha$ among all members. 
The solution for $A_i(t)$ is given by
\begin{equation}\label{eq:Ai}
A_i(t) = -1 + \sqrt{ (m+1)^2 + 2m \ln \left[\frac{m_0+t}{m_0+i}\right]} \ .
\end{equation}
It shows that an older member with smaller $i$ has a larger activity
and that  
the activity grows very slowly in time. 
With the solution for $A$, one can easily show that 
$\sum_{j\leq t}m/(A_j+1)\simeq \Theta_{RV} (m_0 + t )$ for large $t$ with
\begin{equation}\label{eq:Theta}
\Theta_{RV} = \int_0^1 du \frac{m}{\sqrt{ (m+1)^2 -2m \ln u}} \ .
\end{equation}
Hence, the average number of groups increases linearly in time as 
$M(t) \simeq \Theta_{RV} t$ with the group number growth rate $\Theta_{RV}$.
The group size increases algebraically
as 
\begin{equation}\label{eq:S_RV}
S_\alpha(t) \simeq 2 \left( \frac{m_0+t}{m_0+\tau_\alpha} 
\right)^{\Theta_{RV}} \ .
\end{equation}

We have obtained the activity of each member and the size of each group,
which allow us to derive the distribution function $P_a(A)$ and $P_s(S)$
for the activity and the group size, respectively.
The activity distribution function is given by the relation 
$P_a(A) = P_{in}(i) |di/dA|$ 
with the uniform individual distribution, 
$P_{in}(i)= 1/(m_0+t)$.
The differentiation can be done through Eq.~(\ref{eq:Ai}), which
yields that  
the activity distribution is bounded as $P_{a}(A) = (A+1) 
\exp\{ - ( (A+1)^2 - (m+1)^2 )/(2 m) \}/m$. 
Similarly, the group size distribution is given by 
$P_s(S) = P_\alpha(\tau) |d\tau/dS|$
 with the group creation time distribution $P_\alpha(\tau)$.
We assume that the group creation time is distributed uniformly, which is
justified with the linear growth of $M \simeq \Theta_{RV}(m_0+t)$.
Then the group size distribution follows a power law 
$P_s(S) \sim S^{-\gamma_{RV}}$ with the exponent
\begin{equation}
\gamma_{RV} = 1 + \Theta_{RV}^{-1} \ . 
\end{equation}
Note that the distribution exponent is determined by the group number growth 
rate $\Theta_{RV}$.

We now turn to the PF model. With the selection and creation
probabilities,  Eqs.~(\ref{delA},\ref{delM},\ref{delS}) are written as 
\begin{eqnarray}
\frac{d{A}_i}{dt} &=& \frac{m\omega A_i}{\sum_{j\leq t} A_j} 
\label{eq:delA_PF}\\
\frac{dM}{dt} & = & m\omega \label{eq:M_PF}\\
\frac{dS_\alpha}{dt} &=& (1-\omega) S_\alpha \frac{m}{\sum_{j\leq t} A_j} \ .
\end{eqnarray}
We also took the approximation 
$\chi_{i\alpha} = S_\alpha/(m_0+t)$ in Eq.~(\ref{delS}).
Trivially we find that the group number grows in time as 
$M(t) =m\omega t +1$. For $A_i$ and $S_\alpha$, one need evaluate the
quantity $\sum_{j\leq t}A_j$.
Summing over all $i$ both sides of Eq.~(\ref{eq:delA_PF}), one obtains 
that $\sum_{i\leq t}(dA_i/dt) = m\omega$. 
Note that $d({\sum_{i\leq t} A_i})/dt = \sum_{i\leq t} (dA_i/dt)+ m =
(1+\omega)m$, 
which yields that $(\sum_{j\leq t} A_j) = m(1+\omega)t + m_0$. 
Hence we obtain the algebraic growth of the activity and the group size as
\begin{eqnarray}
A_i(t) &=& m \left( \frac{ m(1+\omega)t+m_0 } {
m(1+\omega)i+m_0}\right)^{\frac{\omega}{1+\omega}} \label{eq:A_PF} \\
S_\alpha(t) &=& 2 \left( \frac{ m(1+\omega)t + m_0 }{m(1+\omega)t_\alpha + m_0 }
\right)^\frac{1-\omega}{1+\omega} \label{eq:S_PF} \ .
\end{eqnarray}
These results allow us to find the distribution functions $P_a(A)$ and
$P_s(S)$. They follow the power distribution
$P_a(A) \sim A^{-\lambda_{PF}}$ and $P_s(S) \sim S^{-\gamma_{PF}}$
with the exponents 
\begin{equation}
\lambda_{PF} = 2 + 1 / \omega \quad\mbox{and}\quad \gamma_{PF} =
2/(1-\omega) \ .
\end{equation}
Here we also assumed the uniform distribution of $\tau_\alpha$ in 
Eq.~(\ref{eq:S_PF}), which is supported from the linear growth of $M(t) \sim
m \omega t$.
In contrast to the RV model, both distributions follow the power-law. 
The exponents do not depend on the parameter $m$, but only on the group 
creation probability $\omega$.

For the PV and the RF model, the followings can be
shown easily: The PV model behaves similarly as the RV model.
The group number increases linearly in time
as $M(t) \simeq \Theta_{PV} t$ with the group number growth rate
$\Theta_{PV}$.
Unfortunately, we could not obtain a closed form expression for it. 
However, if we adopt the assumption 
that the selection probability $P_i^S$ is proportional to $A_i+1$ instead
of $A_i$, it can be evaluated analytically as
\begin{equation}\label{eq:Theta_PV}
\Theta_{PV} \simeq \left( \sqrt{m^2+6m+1} - (m+1) \right)/2 \ . 
\end{equation}
The approximation would become better for larger values of $m$.
The group size grows algebraically as in Eq.~(\ref{eq:S_RV}) with
$\Theta_{PV}$ instead of $\Theta_{RV}$. Therefore, the group size
distribution follows the power-law with the exponent $\gamma_{PV}$ presented
in Table~\ref{table1}. The RF model also displays the power-law group size 
distribution. The distribution exponent $\gamma_{RF}$ is given in
Table~\ref{table1}. Note that $\gamma_{RF}$ and $\gamma_{PF}$ are the same.
On the other hand, the activity distribution follows 
an exponential distribution in the RF and the PV model.

Origin for the power-law distribution of the group size is easily
understood. In all models considered, the size of a group increases when
one of its members invites a new member. The larger a group is, the more
chance to invite new members it has. Therefore there exists the 
preferential growth in the group size, which is known to lead to
the power-law distribution~\cite{Barabasi99}.

The activity of a member increases when a newcomer selects it and 
creates a new group.
When the random selection probability is adopted, such a process does not
occur preferentially for members with higher activity. 
It results in the exponential type activity distribution in the RV and 
RF models.
In the PV model, although the selection probability is proportional to the 
activity, the creation probability is inversely proportional to the
activity. Hence, it does not have the preferential growth mechanism in the
member activity either. Only in the PF model, the activity growth rate is
proportional to the activity of each member. Therefore, the activity
distribution follows the power-law only in the PF model.

\begin{figure}[t]
\includegraphics*[width=\columnwidth]{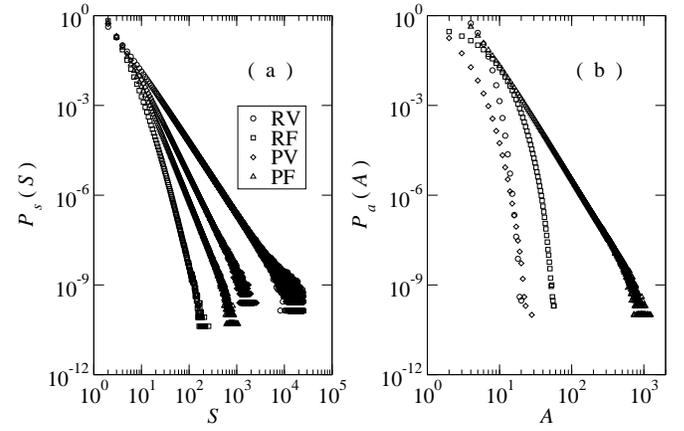}
\caption{(a) The group size distribution and (b) the activity distribution.
The model parameters are $m=4,1$ for the RV and the PV model, respectively.
The RF model has $m=4$ and $\omega=0.6$, and the PF model has $m=4$ and
$\omega=0.5$. The community has grown up to $N=10^6$ and the distributions
are averaged over $10^4$ samples.}
\label{fig3}
\end{figure}
\begin{figure}[t]
\includegraphics*[width=\columnwidth]{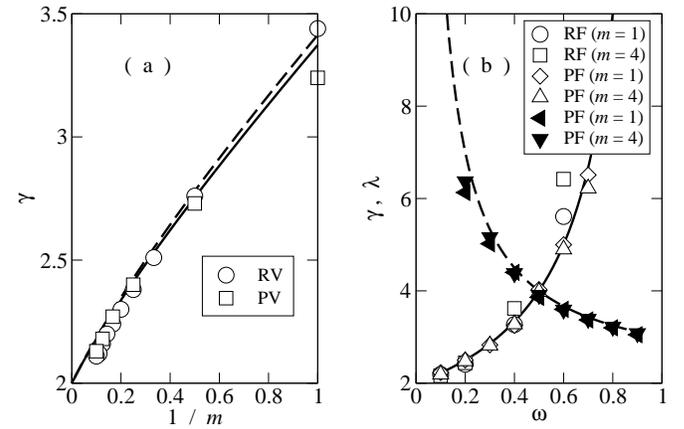}
\caption{(a) Numerical results for $\gamma$ for the RV and the PV model.
The solid~(dashed) curve represents the analytic 
mean field results for the RV~(PV) model. 
(b) Numerical results for $\gamma$~(open symbols) of the RF and the PF
model, and for $\lambda$~(filled symbols) of the PF model.
The solid~(dashed) curve represents the analytic results for
$\gamma$~($\lambda$) in Table~\ref{table1}.}
\label{fig4}
\end{figure}
The analytic mean field results are compared with numerical simulations.
In simulations, we chose $m_0 =m$ and all data were obtained after the
average over at least 10000 samples.
We present the numerical data in Fig.~\ref{fig3}. 
In accordance with the mean field results, 
the group size distribution follows the
power-law in all cases. The activity distribution also shows the expected
behavior; the power-law distribution for the PF model and exponential
type distributions for the other models. 
We summarize the distribution exponents in Fig~.\ref{fig4}.
The measured values of the distribution exponents are in good agreement 
with the analytic results.

Our network models display distinct behaviors from those
bipartite networks such as the movie actor network, 
the scientific collaboration networks, and the director board network 
which have been studied previously. For the first two examples, their
growth is driven only by the member vertices, the actors and the scientists, 
respectively. The activity of members may increase in time. However, the 
group vertices, the movies and the papers, respectively, are frozen
dynamically and their sizes are bounded practically. 
For the last example, both the members~(directors) and the groups~(boards)
may evolve in time. However, it was shown that the group size distribution 
is also bounded~\cite{Newman01a}.

\begin{figure}
\includegraphics*[width=\columnwidth]{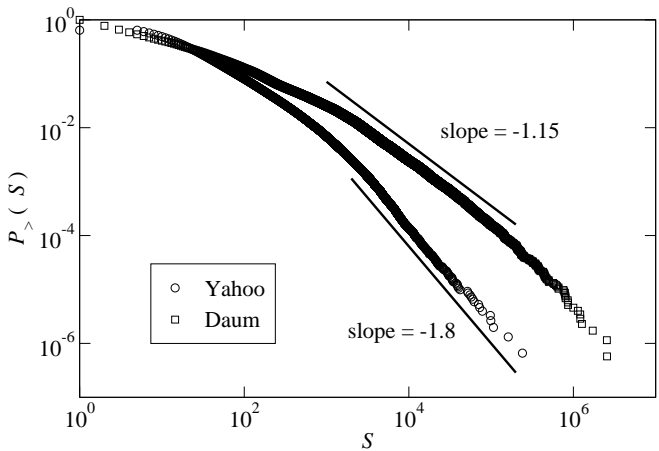}
\caption{Cumulative group size distribution of the online communities in the Yahoo and
the Daum.}
\label{fig5}
\end{figure}

Our model is applicable to evolving networks with the group structure
where the size of a group may increase unlimitedly. 
The online community is a good example of such networks. 
To test the possibility, we study the empirical data obtained from 
the Groups and the Cafe operated by the Yahoo in \url{http://www.yahoo.com}
and the Daum in \url{http://www.daum.net}, respectively. 
It is found in August, 2004 that there are 1,516,750~(1,743,130) groups~(cafes)
with 76,587,494~(351,565,837) cumulative members in the Yahoo~(Daum)
site.
The numbers of members of the groups are available via the web sites.
Figure~\ref{fig5} presents the cumulative distribution
$P_>(S) = \sum_{S' > S} P_s (S')$ of the group size. The distribution has
a fat tail~\cite{comment_data}. Although the distribution function in the log-log scale show a 
nonnegligible curvature in the entire range, it can still 
be fitted reasonable well into the power law for a range 
over two decades~(see the straight lines
drawn in Fig.~\ref{fig5}). From the fitting, we obtain the group size
distribution exponents $\gamma_{\text{Yahoo}} \simeq 2.8$ 
and $\gamma_{\text{Daum}} \simeq 2.15$. The power-law scaling suggests that
the online community may be described by our network model.
Unfortunately, information on the activity distribution is not available
publicly. So we could not compare the activity distribution of the
communities with the model results. 
We would like to add the following remark:
A real-world online community evolves in time as new members are introduced
to and new groups are created. At the same time, it also evolves as members 
leave it and groups are closed. Those processes are not incorporated into
the model. Our model is a minimal model for the online community 
where the effects of leaving members and closed groups are neglected.

\section{Summary}\label{sec:4}
We have introduced the bipartite network model for a growing community with
the group structure. The community consists of members and groups,
gatherings of members. Those ingredients are represented with distinct kinds 
of vertices. And a membership relation is represented with a link between a
member and a group. Upon growing a group increases its size when one of
its members introduces a new member. Hence, a larger group grows
preferentially faster than a smaller group. With the analytic mean field
approaches and the computer simulations, we have shown that the preferential
growth leads to the power-law distribution of the group size.
On the other hand, the activity distribution follows the power-law only for 
the PF model with the preferential selection probability and the fixed 
creation probability~(see Table~\ref{table1}).
We have also studied the empirical data obtained from the online
communities, the Groups of the Yahoo and the Cafe of the Daum. Both
communities display the power-law distribution of the group size. It
suggests our network model be useful in studying their structure.

\acknowledgments{This work was supported by Grant
No. R14-2002-059-01002-0 from the KOSEF-ABRL program and by Grand No. 
KRF-2004--015-C00188.
JDN and HCJ would like to thank KIAS for the support during the visit.}



\begin{references}
\bibitem{comment_Simon} The dynamics of opening a new group or 
joining an existing group is analogous to the copy and growth dynamics 
of the Simon model~(see e.g., Ref.~\cite{Bornholdt01}).
\bibitem{comment_data} The Yahoo groups can be divided into 16 categories and
18,165 sub-categories and the Daum cafes into 22 categories and 825 
sub-categories.
We also find that the group size distribution in each category 
also has the similar fat tail.
\end{references}
\end{document}